# The transition from amorphous to crystalline in Al/Zr multilayers

Qi Zhong, [1] Zhong Zhang, [1] Shuang Ma, [1] Runze Qi, [1] Jia Li, [1] Zhanshan Wang, [1, *]

Karine Le Guen, [2] Jean-Michel André, [2] Philippe Jonnard[2]

[1]MOE Key Laboratory of Advanced Micro-Structured Materials, Institute of Precision Optical Engineering, Department of Physics, Tongji University, Shanghai 200092, China
[2]Laboratoire de Chimie Physique – Matière et Rayonnement, UPMC Univ Paris 06, CNRS UMR 7614, 11 rue Pierre et Marie Curie, F-75231 Paris cedex 05, France

**ABSTRACT:**
The amorphous-to-crystalline transition in Al(1.0%wtSi)/Zr and Al(Pure)/Zr multilayers grown by direct-current magnetron sputtering system has been characterized over a range of Al layer thicknesses (1.0-5.0 nm) by using a series of complementary measurements including grazing incidence X-ray reflectometry, atomic force microscopy, X–ray diffraction and high-resolution transmission electron microscopy. The Al layer thickness transition exhibits the Si doped in Al could not only disfavor the crystallization of Al, but also influence the changing trends of surface roughness and diffraction peak position of phase Al<111>. An interesting feature of the presence of Si in Al layer is that Si could influence the transition process in Al(1%wtSi) layer, in which the critical thickness (1.6 nm) of Al(Pure) layer in Al(Pure)/Zr shifts to 1.8 nm of Al(1.0%wtSi) layer in Al(1.0%wtSi)/Zr multilayer. We also found that the Zr-on-Al interlayer is wider than the Al-on-Zr interlayer in both systems, and the Al layers do not have specific crystal orientation in the directions vertical to the layer from SAED patterns below the thickness (3.0 nm) of Al layers. Above the thickness (3.0 nm) of Al layers, the Al layers are highly oriented in Al<111>, so that the transformation from asymmetrical to symmetrical interlayers can be observed. Based on the analysis of all measurements, we build up a model with four steps, which could explain the Al layer thickness transition process in terms of a critical thickness for the nucleation of Al(Pure) and Al(1%wtSi) crystallites.

## 1. INTRODUCTION

The crystallization in the multilayers can influence the performances of the multilayers, such as optical properties,[1-4] electrochemical effects [5-7] and thermal reactions.[8-11] In Al/Zr multilayers, the inhomogeneous crystallization of Al and interdiffusion between Al and Zr are the phenomena which affect the reflectance of the multilayers. Although the theoretical reflectance of Al(1.0%wtSi)/Zr multilayers with 40 periods is 70.9% at 5 °incidence angle, the experimental reflectance is only 41.2%.[2, 3] Based on the simulation of EUV reflectance, four factors responsible for the loss of reflectance are the inhomogeneous crystallization of Al, contamination of the multilayer, surface oxidized layer and interdiffusion between Al and Zr layers. To improve the experimental reflectance, we should overcome these four limitations in the further experiments. In particular, the optical and structural performances of Al(1.0%wtSi)/Zr multilayers are mainly influenced by the crystallization of Al**.** For large period numbers of Al(1.0%wtSi)/Zr multilayers, the interfacial roughness in the multilayers varies. Indeed, with less than 40 periods, the roughness components are smoothened by the multilayers. But for more than 40 periods, the surface and interfacial roughness are accumulated with the increasing period number. For these multilayers, it is found that the Al is in a polycrystalline fcc phase with a highly preferred <111> texture, and the Zr is in an hcp phase. The images of high-resolution transmission electron microscopy (TEM) and X-ray diffraction (XRD) data also reveal that the variable interfacial and surface roughnesses are mainly caused by the inhomogeneous crystallization of Al, which grain size is equal to the thickness of Al layer. In order to reduce the influence of the inhomogeneous crystallization of Al on the optical and structural performances, we should first know the critical thickness of an amorphous-to-crystalline transition in the Al layers. Then, based on that critical thickness value, investigations will focus on the interface formation and the optimization of the multilayer structure with respect to smooth interfaces and increase the EUV reflectance. However, there is no data found on the critical thicknesses of the transition region in Al(1.0%wtSi)/Zr and Al(Pure)/Zr multilayers.

In this paper, we present the Al layer thickness transition from amorphous to crystalline in both Al(1.0%wtSi)/Zr and Al(Pure)/Zr multilayers. The purpose is to investigate how the presence of silicon in the Al layers influences this transition, and the structural performance of the thickness transition. The deposition processes of multilayers are outlined in Sec.2. The critical thicknesses of the amorphous-to-crystalline transition in Al(1.0%wtSi)/Zr and Al(Pure)/Zr multilayers are characterized by using grazing incidence X-ray reflectometry (GIXR), XRD, atomic force microscopy (AFM) and TEM separately in Sec. 3. We also present an explanation for the different thickness transition processes in the two systems (Sec. 3). In Sec.4, we conclude with the performances and comparison of transition thicknesses for Al(1.0%wtSi)/Zr and Al(Pure)/Zr systems.

## 2. EXPERIMENT

All Al/Zr multilayers were prepared by using the direct-current magnetron sputtering system. The base pressure was $8.0 \times 10^{-5}$ Pa, and the samples were deposited on Si polished wafers under a 0.16 Pa argon (99.9999% purity) pressure. The sputtering targets with diameter of 100 mm were zirconium (99.5%), aluminum (99.999%, Al(Pure)) and silicon doped in aluminum (Al(1%wtSi)). In the different multilayers, the Zr layer thicknesses are held constant around 4.0 nm while the Al layer thicknesses are varied from 1.0 to 5.0 nm. To keep same total thickness of the stack, the number of periods is variable as shown in Table 1.

Table 1. Thicknesses of Al and Zr layers in Al(1%wtSi)/Zr and Al(Pure)/Zr multilayers with different periods as derived from the GIXR simulation.

|  | Sample No. | 1 | 2 | 3 | 4 | 5 | 6 | 7 | 8 |
|---|---|---|---|---|---|---|---|---|---|
| **Al(1%wtSi)/Zr** | Al(1%wtSi) layer/nm | 1.0 | 1.6 | 1.7 | 1.8 | 2.0 | 3.2 | 4.1 | 4.8 |
|  | Zr-on-Al/nm | 1.0 | 1.0 | 1.0 | 1.0 | 1.0 | 1.5 | 1.5 | 1.5 |
|  | Al-on-Zr/nm | 0.6 | 0.6 | 0.6 | 0.6 | 0.6 | 1.5 | 1.5 | 1.5 |
|  | Zr layer/nm | 4.1 | 4.2 | 4.2 | 4.2 | 3.9 | 3.8 | 3.9 | 4.1 |
|  | Periods (N) | 60 | 40 | 36 | 33 | 30 | 20 | 15 | 12 |
|  | Sample No. | 9 | 10 | 11 | 12 | 13 | 14 | 15 |  |
| **Al(Pure)/Zr** | Al(Pure) layer/nm | 0.8 | 1.6 | 1.7 | 2.1 | 3.1 | 4.2 | 5.1 |  |
|  | Zr-on-Al/nm | 1.0 | 1.0 | 1.0 | 1.0 | 1.5 | 1.5 | 1.5 |  |
|  | Al-on-Zr/nm | 0.6 | 0.6 | 0.6 | 0.6 | 1.5 | 1.5 | 1.5 |  |
|  | Zr layer/nm | 4.1 | 3.9 | 3.9 | 3.9 | 4.0 | 3.6 | 3.8 |  |
|  | Periods (N) | 60 | 40 | 36 | 30 | 20 | 15 | 12 |  |

The values extracted from GIXR measurements shown in Table 1 were obtained by using an x-ray diffractometer working at the Cu $K_\alpha$ line (0.154 nm). The fitting calculations of GIXR curves were performed with Bede Refs software (genetic algorithm) to determine individual layer thickness and interface roughness [12]. The XRD measurements provide identification of crystalline phases and estimation of crystal size. While the surface roughness was measured with a Veeco, Multi-Mode SPM scanning probe microscope, operated in AFM mode. The uncertainty on the determination of surface roughness is that only one point could not represent the whole surface area. Therefore, we measured three points from different parts in one sample. And the error can be reduced to 0.05nm. To further confirm the Al layer thickness transition from amorphous to crystalline in both Al/Zr systems, the transmission electron microscope (TEM, FEI Tecnai G$^2$ F20) was used on the specimen prepared by focused ion beam (FIB, Materials Analysis Technology Ltd.).

## 3. RESULTS AND DISCUSSION

### 3.1 Grazing incident X–ray reflection

To estimate the individual layer thicknesses in the Al/Zr (Al(1%wtSi)/Zr and Al(Pure)/Zr) multilayers, all samples were characterized by GIXR, measured over the angular range of θ=0 °~4 °. The GIXR data is fitted by four-layer asymmetrical and symmetrical models when the thickness of Al layers is below and above 3.0 nm, respectively. The GIXR produces a series of sharp peaks corresponding to diffraction from the multilayer structure. The examples of the GIXR spectra and fitting data from the Al(1%wtSi)/Zr (No. 3) and Al(Pure)/Zr (No. 10 in Table 1) are shown in Figure 1. From the fitting data (not all fitting data presented in Figure 1), the thicknesses of all layers, especially for the Al layers, are presented in the Table 1. The Zr layer is kept around 4.0 nm. The Al layers for Al(1%wtSi)/Zr and Al(Pure)/Zr multilayers vary from 1.0 to 4.8 nm, and 0.8 to 5.1 nm, respectively. In both systems the thickness (1.0 nm) of Zr-on-Al interlayers are thicker than that (0.6 nm) of Al-on-Zr interlayers when the thickness of Al layers is smaller than 3.0 nm. This is due to the Al(fcc) crystal is not obviously in the Al layers, which the free energies of the formation between Al and Zr interfaces are not balanced to form the symmetrical interlayers. However, the interlayers become symmetrical above 3.0 nm of Al layer, when the phase Al<111> of Al-fcc is highly oriented in the multilayers. Because the different periods could influence the roughness, the interfacial roughnesses of all multilayers are not compared in the Table 1.[2]

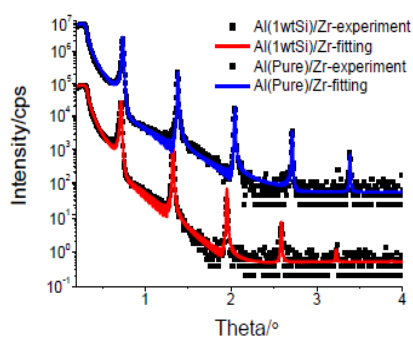

Figure 1. Comparison between the GIXR experimental (black dots) and fitting (color lines) curves for the Al(1%wtSi)/Zr (No. 3) and Al(Pure)/Zr (No.10)

### 3.2 Atomic force microscopy

The surface roughnesses of the Al(1%wtSi)/Zr and Al(Pure)/Zr multilayers are revealed by AFM, measured over a 5 μm × 5 μm area. In Figure 2, we show the surface roughness of some samples below the Al layer thickness transition (d$_{Al(1\%wtSi)}$=1.0 and 1.6 nm, d$_{Al(Pure)}$=0.8 nm, within the Al layer thickness transition (d$_{Al(1\%wtSi)}$= 1.7, 1.8 and 2.0 nm, d$_{Al(Pure)}$=1.6 nm, and above the Al layer thickness transition ($d_{Al(1\%wtSi)}$=4.1 and 4.8 nm, $d_{Al(Pure)}$=1.7, 2.1, 4.2 and 5.1nm). For the Al(Pure)/Zr multilayers, the surface has sparsely packed small particles below the transition and a roughness of 0.19 nm. Within the transition, the roughness has a significant peak height of 0.50 nm. Above the transition, the surface roughnesses are about 0.30 nm. Comparing Al(1%wtSi)/Zr and Al(Pure)/Zr multilayers in Figure 2, it is observed that the changing trends in the two systems present different situations. For Al(1%wtSi)/Zr multilayers, the surface roughness first slightly increases below the transition. The roughness is 0.21 nm at the thickness (1.6 nm) of Al(1%wtSi), in which the surface consists of sparsely packed small particles is very smooth. In particular, the amorphous-to-crystalline transition can be identified by a sharp, transient increase in the roughness. [13] The roughness increases from a value of 0.21 nm below the transition to a peak value of 0.57 nm within the transition at the thickness (1.8 nm) of Al(1%wtSi) layer. Above the transition, the surface roughness begins to decrease, but it is still much higher than the value below the transition. An example of the roughness at the thickness (4.8 nm) of Al(1%wtSi) layer is 0.45 nm. For the Al(Pure)/Zr multilayers, considering the error (±0.05nm) of the surface value, we can found that the surface roughnesses are similar below and above the transition, which values are around 0.20 and 0.30 nm, respectively, but significantly increases within the transition region. An interesting feature in the Figure 2 is that the surface roughness above the transition in Al(Pure) layer is smaller than that in Al(1%wtSi) layer which may be influenced by the crystallization of the different Al layers. Based on the analysis of AFM, we can assume that the crystal growth processes in Al layers are much different in two systems.

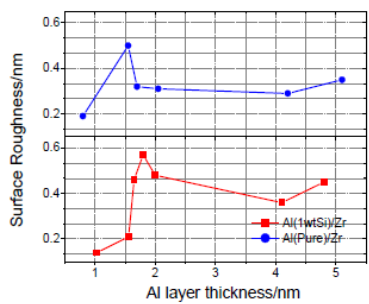

Figure 2. Surface roughness as a function of the thicknesses of Al layers for two Al/Zr (Al(1%wtSi)/Zr and Al(Pure)/Zr) systems.

### 3.3 X–ray diffraction

To verify the AFM analysis in previous section, the XRD measurements are presented in Figure 3. In order to ignore the influence of the different thicknesses of Zr layers, all XRD data are normalized by the highest intensity in the corresponding curve. From Figure 3.a and b, we can see the phase of Al<111> at the thicknesses (4.8 nm for Al(1%wtSi)/Zr and 5.1 nm for Al(Pure)/Zr), which has two different positions 38.5 °[14] and 38.7 °[15], respectively. This may be influenced by the different properties and densities of the Al materials in the multilayers. In Figure 3.a, below the transition, there is one amorphous peak around 38.0 °. But within and above the transition, the peak position becomes closer to 38.5 ° with the increasing thickness of Al(1%wtSi) layers. In Al(Pure)/Zr multilayers (Figure 3.b), we can also find an amorphous peak below the transition. Its position shifts from 38.1 ° to 38.7 ° over the considered thickness range.

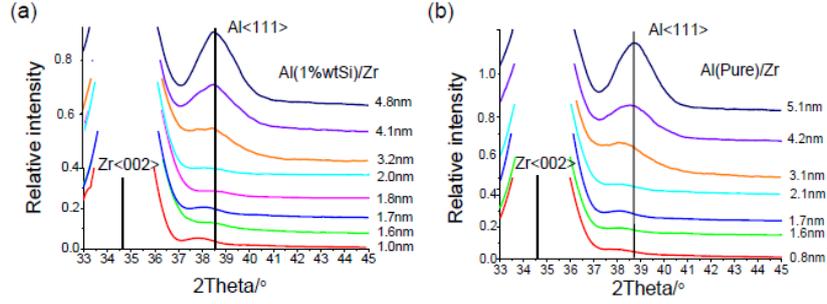

Figure 3. XRD measurements relative intensity vs 2theta angle of for Al(1%wtSi)/Zr (a) and Al(Pure)/Zr (b) multilayers. The samples with variable thickness of Al layers are in different colors.

We compared in Figure 4 the diffraction peak position and crystal size of Al(1%wtSi)/Zr and Al(Pure)/Zr multilayers as a function of the Al layer thickness. In Figure 4.a, there are three changing regions in Al(1%wtSi)/Zr systems. The first region is from 1.0 to 1.6 nm, and then the second region is from 1.6 to 3.2 nm. The last region is from 3.2 to 4.8 nm. The different changing trends in the different regions may be influenced by the crystallization of Al in the different Al layer thicknesses, which is consistent with the three regions determined by AFM in Figure 2. After the thickness increases to 3.2 nm, the diffraction peak position has a constant value at $(38.5 \pm 0.05)°$. But for Al(Pure)/Zr multilayers, the diffraction peak position is at 37.8° below the transition. Within the transition, the slope of the curve is increased. Above the transition, when the thickness of Al(Pure) layers is below 3.0 nm, the diffraction peak position changes slightly. Above the thickness of 3.0 nm for Al(Pure) layers, the diffraction peak position obviously increases from 38.2° to 38.7°. As we know, the surface roughness is influenced by the crystallization of Al layer. The different diffraction peak positions of Al<111> in the two kind of multilayers mean that the corresponding crystal lattices are formed differently in the two systems, the surface roughness in Al(1%wtSi)/Zr multilayers being more sensitive to the crystallization of Al layers than that in Al(Pure)/Zr multilayers. This could explain the situation that the surface roughness of Al(1%wtSi)/Zr is higher than that of Al(Pure)/Zr in Figure 2.

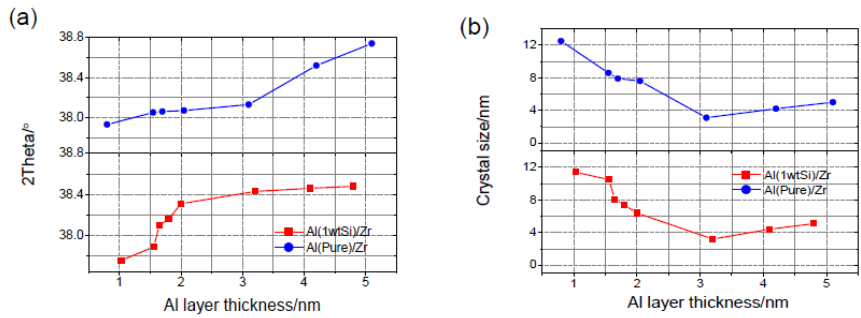

Figure 4. Changing trends of 2theta (a) and crystal size (b) with increasing Al layer thicknesses for the two Al/Zr systems.

Considering the Scherrer formula [15], we derive in Figure 4.b the crystal size of Al crystallites from Figure 3. The crystal sizes of two systems have three areas. First, there are much larger size (>10.0 nm) of Al(1%wtSi) layers from 1.0 to 1.6 nm, and Al(Pure) layers at 0.8 nm, respectively. The thickness from 1.7 to 2.0 nm for Al(1%wtSi) layers and 1.6 to 2.1 nm for Al(Pure) layers correspond to the second region. The crystal sizes in this region become smaller, but remain larger than the thickness of Al layer. Finally, above a 3.0 nm thickness of Al(1%wtSi) and Al(Pure) layers, the crystal size is almost equal to the thickness of Al layers. As an example of the 4.1 nm thickness of Al(1%wtSi) layer corresponds to a crystal size 4.2 nm.

Based on the results in GIXR, AFM and XRD, we can deduce that the transition from amorphous to crystalline states appears in both Al(1%wtSi)/Zr and Al(Pure)/Zr systems. Below the transition, the Al layers are all amorphous. Within the transition, the crystallites of Al can nucleate, but some parts of the Al layers are still amorphous. Above the transition, but for Al thickness lower than 3.0 nm, the Al crystallites are not highly oriented in the multilayers and cannot greatly influence the interfacial structure. Above the 3.0 nm thickness, Al crystallites are highly oriented and influence the interfacial structure (Table 1), and Al<111> presents remarkable peak in the XRD measurements (Figure 3). The changing trends of the XRD measurements are much different in the two systems. The Si introduced in Al layers disfavors the crystallization of Al and shifts the position of the Al<111> diffraction peak. To sum up, we believe that the Al layers have different growth processes in Al(1%wtSi)/Zr and Al(Pure)/Zr systems.

### 3.4 High-resolution transmission electron microscopy

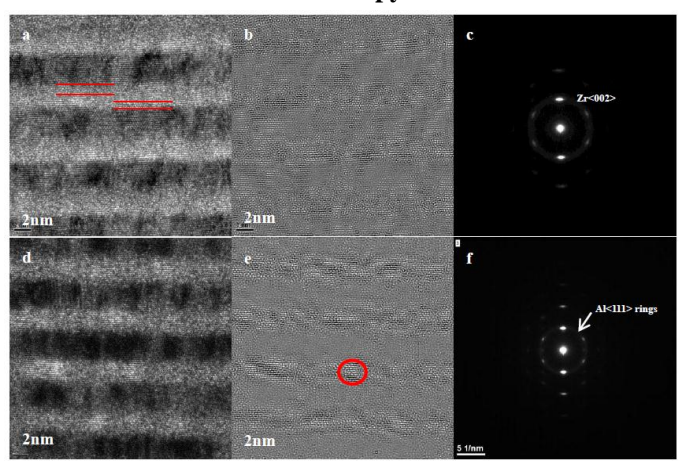

Figure 5. Cross-sectional TEM images of Al(1%wtSi)/Zr multilayer samples in bright field (a, d), Fourier-filtering corresponding to a and d images (b and e) and SAED patterns (e, f) for the samples of Al(1%wtSi)/Zr (Al=1.6 nm, a, b, c) and Al(1%wtSi)/Zr (Al=1.7 nm, d, e, f)

To further confirm the transition from amorphous to crystalline states of the Al layers, TEM measurements are carried out on Al(1%wtSi)/Zr multilayers. From the previous sections, we can get that the changing trends of surface roughness, diffraction peak position and crystal size in Al(1%wtSi) and Al(Pure) layers are similar below and within the transition. Therefore, we can use the images of Al(1%wtSi)/Zr to represent the difference between below and within the transition in both systems. The images correspond to Al(1%wtSi)/Zr samples having Al(1%wtSi) layer thicknesses below (Figure 5.a) and within (Figure 5.d) the transition. The dark layers are Zr and the light layers are Al(1%wtSi). In order to enhance the contrast of Al crystal grains, the image processing by a Fourier-filtering of TEM image of Figures 5.a and d is shown in Figures 5.b and e. Below the transition, Al layers are mostly amorphous. Although we can find some Al grains, the crystallizations of Al is not well formed, which present a smooth interface and a low surface roughness (Figure 2). In the selected area electron diffraction (SAED) pattern, Figure 5.c, the Al(1%wtSi) grains do not have specific orientation in relation with the direction perpendicular to the layers and just present the phase of Zr layers. Within the transition, the Al(1%wtSi) layer is beginning to nucleate, and the grains of Al(1%wtSi) are almost intact at few points, which could influence the microstructure of the multilayers (Figure 5.e). But some parts of Al(1%wtSi) layer are still amorphous, due to the presence of Si in Al layer and large interdiffusion between Al and Zr layers. Because of the nucleation in Al(1%wtSi) layers, the diffraction peak position and surface roughness are changed, consistent with the XRD (Figure 4.a) and AFM (Figure 2) results. The SAED pattern (Figure 5.f) also reveals that the diffraction pattern of Al(1%wtSi) layer is more obvious than that for below the transition in Figure 5.c and confirms that the transition from amorphous to crystalline appears at the thickness of 1.7 nm for the Al(1%wtSi) layers.

Considering the interdiffusion, we found the interlayers in both samples are asymmetrical with the Zr-on-Al interlayer wider than the Al-on-Zr interlayer. An example (Figure 5.a) for of Al(1%wtSi) 1.6 nm thick layers, the thickness of Zr-on-Al interlayer is 1.1 nm, and that of Al-on-Zr interlayer is 0.6 nm, which are consistent with the fitting data in GIXR (Table 1). The reason of the asymmetrical interlayers is that the Al<111> phase is not highly oriented when the thickness of Al(1%wtSi) layers is smaller than 3.0 nm. Above 3.0 nm, the interlayers become symmetrical; both interlayers are having a 1.5 nm thickness.[11]

### 3.5 Discussion

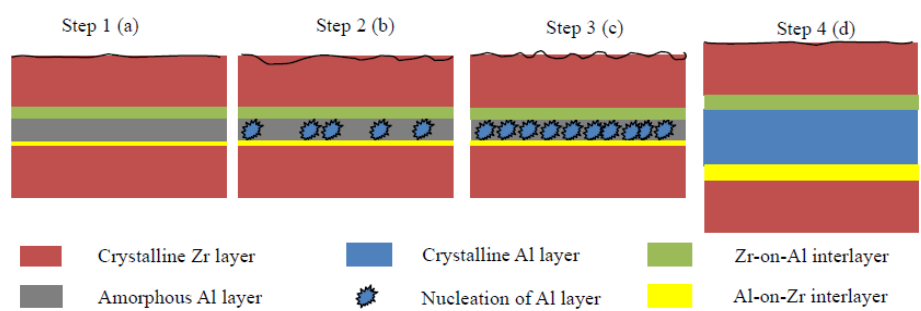

Figure 6. A few simplified pictures of the amorphous-to-crystalline transition in Al layers. (a) Below the critical thickness, the Al layer is entirely amorphous and the Zr-on-Al interlayer is wider than the Al-on-Zr one. (b) As the Al layer grows within the transition but does not exceed the critical thickness, just a few isolated nucleation points appear, especially for the Al(1%wtSi)/Zr systems. The Zr-on-Al interlayer is still the thickest one. (c) When the thickness of Al layer just at the critical thickness, Al crystallites nucleate in most points of the layer, which increases the interfacial and surface roughness. (d) Above the critical thickness the Al layer is mostly polycrystalline, there is less roughness and the interlayers are symmetrical.

To explain the results from GIXR, AFM, XRD and TEM measurements, we present a growth model for the transition in the Al(1%wtSi) and Al(Pure) layers that is based on the concept of critical thickness [13] for the nucleation of a crystalline phase. The amorphous-to-crystalline transition exhibits several interesting features:

(1) The surface roughness increases with the Al layer thickness in both systems and has the highest value at the transition. However, below and above the transition, the changes of the roughness are different in Al(1%wtSi)/Zr and Al(Pure)/Zr multilayers (Figure 2).
(2) The diffraction peak positions of Al<111> have different changing trends in the two systems.
(3) Below a 3.0 nm thickness, the Al layer is not oriented. The Zr-on-Al interlayer is wider than the Al-on-Zr interlayer. For Al layers thicker than 3.0 nm, the size of the Al crystallites in the growth direction is equal to the thickness of the Al layers and the interlayers become symmetrical.

To interpret these observations, we present a simple model with four steps for the transition of the Al layers. Considering the growth direction of Zr layers (in-plane), the Al grains (out-plane) could not only influence the surface roughness, but also impact the growth of Zr grains and the symmetry of interface. Using the concept of the critical thickness in the Mo/Si multilayers [13], characteristics of the amorphous-to-crystalline transition should follow naturally. Our four-step model for the crystallization of the Al layers is shown schematically in Figure 6. Based on the results in the previous sections, the performances of Al(1%wtSi)/Zr and Al(Pure)/Zr are different and should be described separately. For the Al(Pure)/Zr system, when the Al(Pure) layer thickness is below the critical thickness (1.6 nm), as shown in Figure 6.a (Step-1), the Al(Pure) layer is amorphous. Because the Al(Pure) layer is smooth, and the growth direction of Zr layer is in-plane, the Zr layer deposited on top of the Al(Pure) layer is also smooth, which results in a smooth surface of the samples. When the average thickness of the layer is just at the critical thickness (Figure 6.c, Step-3), and the critical thickness will first be reached at the extreme points, in which the crystallites of Al(Pure) can nucleate at those points. The nucleation of Al(Pure) crystallites affect the growth of the overlaying Zr layer leading to a rougher surface. Because the Al<111> is not highly oriented when the thickness of Al(Pure) layers is below 3.0 nm, the free energies of Al(Pure) and Zr layers are different. Therefore, Zr diffuses and reacts more rapidly with Al than Al diffuses and reacts with Zr [16], producing asymmetrical interlayers in the multilayers. With the increasing Al(Pure) layer thickness above the transition region, especially above 3.0 nm (Figure 6.d, Step-4), the layer is mostly crystalline. The phase of Al<111> is highly oriented in the samples and the diffraction peak position of Al<111> is close to 38.7 °. The grain size in the growth direction is equal to the thickness of Al(Pure) layer, leading to a smooth surface in the multilayers. At present, the free energies between Al(fcc) and Zr(hcp) layers are equal [17], so that the interface becomes symmetrical. This kind of growth process is similar to the model for the Mo/a-Si multilayers [13], except the symmetrical transformation in the interlayers.

For Al(1%wtSi)/Zr multilayer, below the critical thickness, 1.8 nm, the situation is the same as with Al(Pure)/Zr systems (Figure 6.a, Step-1). For example, the crystallization of Al(1%wtSi) cannot nucleate at thickness of 1.6 nm (as shown in Figure 5.b). Indeed, the presence of Si in Al layers disfavors the nucleation of Al(1%wtSi). In order to nucleate, the thickness should be thicker. Within the transition, but when the average thickness, 1.7 nm, is below the critical thickness (Figure 6.b, Step-2), the crystallization of Al(1%wtSi) can begin at some points, consistent with the image in Figure 5.e. But most of the layers are still amorphous, which changes the surface roughness slightly. At the critical thickness (Figure 6.c, Step-3), the nucleation of Al(1%wtSi) reaches the most extreme points, which the grains of Al(1%wtSi) are well formed at those points. The corresponding surface roughness is at its maximum (Figure 2). Above the critical thickness, but when the average thickness of Al(1%wtSi) is not over 3.0 nm, the interlayers still present an asymmetry. Above the 3.0 nm thickness (Figure 6.d, Step-4), the Al(1%wtSi) layer becomes fully crystalline, but due to the presence of Si in Al(1%wtSi), the diffraction peak position of Al<111> remains constant at 38.5 °(Figure 4.a).

To sum up, the four-step model can explain the GIXR, AFM, XRD and TEM results. The critical thickness, 1.8 nm, of Al(1%wtSi) layers in Al(1%wtSi)/Zr is larger than that, 1.6 nm, of Al(Pure) layers in Al(Pure)/Zr multilayers. Moreover, the interlayers transform from asymmetrical to symmetrical at a thickness of 3.0 nm in both systems.

## 4. CONCLUSION

We observe that as the thickness of Al layers in Al(Pure)/Zr and Al(1%wtSi)/Zr systems is increased, the Al layers exhibit a transition from amorphous to crystalline at the thickness of 1.6 nm and 1.8 nm, respectively. The presence of Si in Al could not only disfavors the crystallization of Al layers, but also influence the Al<111> diffraction peak position and finally degrades the surface roughness. Our model explains the GIXR, AFM, XRD and TEM observations and the different transition behaviors during the growth processes of Al(Pure)/Zr and Al(1%wtSi)/Zr multilayers. Below the critical thickness (1.6 nm for Al(Pure) layers and 1.8 nm for Al(1%wtSi) layers), the Al layers are amorphous with the lowest surface roughness and smooth interfaces. The Zr-on-Al interlayer is thicker than the Al-on-Zr one. At critical thicknesses, the surface roughness is at maximum. Above the critical thickness, but below 3.0 nm, the Al crystallites are not highly oriented and the interfaces are still asymmetrical. Above the thickness of 3.0 nm, the changing trends of diffraction peak positions in Al(Pure)/Zr and Al(1%wtSi)/Zr multilayers are different, leading to the higher surface roughness in Al(1%wtSi)/Zr multilayers. The Al<111> phase is highly oriented and the grain size is equal to the thickness of Al layer. The interlayers transform from asymmetrical to symmetrical in the multilayers. Based on this explanation, we have a clear understanding of the transition from amorphous-to-crystalline in Al(Pure)/Zr and Al(1%wtSi)/Zr systems. Using the critical thicknesses of Al layers, we will be able to solve the problem of inhomogeneous crystallization of Al layers and will optimize the multilayer structure in the future applications.

■ **AUTHOR INFORMATION**

**Corresponding Author**
* Tel.: +86-021-65984652; fax: +86-021-65984652.
  E-mail: wangzs@tongji.edu.cn

**Notes**
The authors declare no competing financial interest.

■ **ACKNOWLEDGEMENTS**

This work is supported by National Basic Research Program of China (No. 2011CB922203) and National Natural Science Foundation of China (No.10825521, 11027507). Part of this work was done in the framework of the COBMUL project funded by both Agence National de la Recherche in France (#10-INTB-902-01) and Natural Science Foundation of China (#11061130549).